\documentclass[dvips]{acta}
\usepackage{supertabular,lscape,epsfig}
\usepackage{amssymb}
\usepackage{amsmath}

\SetPages{0}{0}

\SetVol{0}{0}

\def\st2{St~2$-$22}
\def\kms{km\,s$^\mathrm{-1}$}
\def\fs{\ensuremath{\overset{\text{s}}{.}}}
\def\degr{\ensuremath{^\circ}}
\def\farcs{\ensuremath{\overset{\prime\prime}{.}}}
\def\arcmin{\ensuremath{^\prime}}

\def\sun{\ensuremath{\odot}}
\def\fm{\ensuremath{\overset{\text{m}}{.}}}

     \usepackage{xcolor}  \definecolor{myred}{RGB}{166,0,26} 

\begin{document} 
	
\begin{Titlepage}
\Title{\st2\ -- another symbiotic star with high-velocity bipolar jets}

\Author{T~o~m~o~v, T.,}{Centre for Astronomy, Faculty of Physics, Astronomy and Informatics, Nicolaus Copernicus University, Grudziadzka 5, 87-100 Torun, Poland\\
			e-mail:tomov@umk.pl}
		
\Author{Z~a~m~a~n~o~v, R.,}{Institute of Astronomy and National Astronomical Observatory, Bulgarian Academy of Sciences, Tsarigradsko Shose 72, 1784 Sofia, Bulgaria\\
	e-mail:rkz@astro.bas.bg}	

\Author{G~a~\l~a~n, C.,}{Nicolaus Copernicus Astronomical Center, Polish Academy of Sciences, Bartycka 18, PL-00-716 Warsaw, Poland\\
	e-mail:cgalan@camk.edu.pl}	

\Author{P~i~e~t~r~u~k~o~w~i~c~z, P.,}{Warsaw University Observatory, Al. Ujazdowskie 4, 00-478 Warszawa, Poland\\
	e-mail:pietruk@astrouw.edu.pl}

\Received{Month Day, Year}

\end{Titlepage}

\Abstract{We report finding high-velocity components in the H$\alpha$ emission wings of \st2\ spectra, obtained in 2005. This discovery have encouraged us to start the present study, aiming to show that this little studied object is a jet producing symbiotic system. We used high-resolution optical and low-resolution near infrared spectra, as well as available  optical and infrared photometry, to evaluate some of the physical parameters of the \st2\ components and the characteristics of the jets. The evaluated parameters of the components confirmed that \st2\ is a S-type symbiotic star. Our results demonstrate that an unnoticed outburst of \st2, similar to those in classical symbiotics, occurred in the first half of 2005. During the outburst, collimated, bipolar jets were ejected by the hot component of \st2 with an average velocity of about 1700\,\kms.}
{stars: binaries: symbiotic -- individual: \st2\ -- interstellar medium: jets and outflows}

\section{Introduction}

Collimated jets have been observed in many types of astrophysical objects -- from pre-main sequence stars to active galactic nuclei (Livio 1999, 2011). Recently, it became clear that the most powerful jets are likely related to the gamma-ray bursts (Granot \& van der Horst 2014). It is generally accepted that the existence of an accretion disk around the central object is a common feature for all jet-producing systems. According to Livio (1999, 2011) the jets acceleration and collimation mechanisms are the same in all classes of astrophysical objects which produce jets and the production of powerful jets requires an additional heat/wind source 
associated with the central object.

Symbiotic stars are wide binary systems with orbital periods of the order of years. They consist of a red giant and compact component which accretes matter from the cool giant's wind. In nearly all systems the compact component is a white dwarf (Kenyon 1986). The symbiotic stars belong to the astrophysical objects in which high-velocity bipolar outflows are not uncommon. Many symbiotics are relatively bright and can be studied in detail with middle class telescopes, what makes them one of the most promising targets for studying jets. Such investigations will shed more light on the processes of ejection, collimation and acceleration of jets not only for systems with white dwarf as central objects but for all jet-producing systems. The recent high-resolution observations of the central part of the R~Aqr jets are good example (Schmid \etal 2017).

Until now, for about a dozen among the more than 200 symbiotic stars, high-velocity bipolar jets has been observed. These outflows are detectable by imaging and spectroscopy in a wide spectral region from X-ray to radio (\eg Taylor \etal 1986; Tomov \etal 1990; Karovska \etal 2007; Angeloni \etal 2011). In some objects \eg Hen~3$-$1341 and Z~And the jets are transient and appear during outburst only (Tomov \etal 2000; Munari \etal 2005; Skopal \etal 2009) while in others like MWC~560 they present permanently (Tomov \& Kolev 1997; Schmid \etal 2001). Because of the intrinsic, long-term variability of the symbiotic stars it is very difficult to carry out a systematic monitoring in the search for high-velocity bipolar jets. Therefore, most of them have been discovered by chance.

In this paper, we report on such a detection of optical, high-velocity, collimated, bipolar jets in the southern symbiotic system \st2. It is a very poorly studied object, included in the Allen (1984) and Belczy\'{n}ski \etal (2000) catalogues of symbiotic stars as \st2\ but shown in SIMBAD as PN~Sa~3$-$22 ($\alpha_\mathrm{2000}=13^\mathrm{h}14^\mathrm{m}30\fs30, \delta_\mathrm{2000}=-58\degr51\arcmin49\farcs59$). Discovered by Sanduleak (1976) \st2\ was classified as planetary nebula, which caused a long-term confusion concerning the nature of the object. Allen (1984) identified the Raman scattered line 6825\,\AA\ in its spectrum, which gave him an undoubted argument to include \st2\ in the catalogue of symbiotic stars. The data for \st2\ in the literature is very scarce. There is no information on observed outbursts of the star so far. Van Winckel \etal (1993) have observed H$\alpha$ in 1988 and 1992 and reported that "the object underwent marked brightness variations in recent years" but gave no details about the nature of these variations. Garc\'{i}a \etal (2003) observed the system while searching linear polarization but the detection was negative. In the paper of Zamanov \etal (2008), \st2\ is included among the symbiotics for which the rotational velocities of the giants were studied. M\"{u}rset \& Schmid (1999) have estimated a spectral class M4.5 for the giant in the system. Miko\l{}ajewska \etal (1997) have derived the reddening $E(B-V)\sim1^\mathrm{m}$ and distance $d\sim5$\,kpc\ of the system and the temperature $T_\mathrm{h}\sim54\div100\times10^\mathrm{3}$\,K and luminosity $L_\mathrm{h}\sim600$\,L$_\sun$ of the hot component.


\MakeTable{llllll}{12.5cm}{Journal of spectroscopic observations.}
{\hline
Date & UT  & Wavelength & R & Exposure & Instrument  \\ 
& middle & range & $\lambda/\Delta\lambda$ & [sec] &  \\
\hline 
01.02.2005 & 06:16:41 & 3600 -- 9200\,\AA & 48000 & 2$\times$1800 & FEROS \\ 
16.05.2005 & 02:04:47 & 3600 -- 9200\,\AA & 48000 & 2$\times$1800 & FEROS \\ 
18.06.2016 & 23:38:36 & 0.94 -- 1.64\,$\mu$m & 600 -- 700 & 4$\times$80 & SOFI \\ 
18.06.2016 & 23:48:05 & 1.50 -- 2.53\,$\mu$m & 600 -- 900 & 4$\times$70 & SOFI \\ 
\hline 
}

\section{Observations and data reduction}

We used spectra of \st2\  obtained with FEROS at the 2.2-m telescope (ESO, La Silla), under programme 074.D-0114, on 2005 February 1st and May 16th. FEROS is a fibre-fed echelle spectrograph, providing a high resolution of 48000, a wide wavelength coverage from about 3600\,\AA\ to 9200\,\AA\ in one exposure and a high efficiency (Kaufer \etal 1999). The 39 orders of the echelle spectrum are registered on a 2k$\times$4k EEV CCD. The spectra were reduced using the dedicated FEROS data reduction software implemented in the ESO-MIDAS system. The achieved S/N ratio in the region of H$\alpha$ is $\sim30$. The spectra were reduced to fluxes by the  use of the spectrophotometric standards HR~3454 and HR~4963, observed  at similar zenith distances during the first and the second nights respectively. The fluxes were de-reddened with $E(B-V)=1\fm0$ (see Section~3.2) and using the standard interstellar medium extinction curve of Fitzpatrick (1999). As an example, the spectral region between 4625\AA\ and 5025\AA\ is shown in Fig.~1.

The near infrared spectra of \st2\ were obtained with SOFI spectrograph on the ESO NTT telescope in low-resolution mode using Blue and Red grisms. The observations were acquired on 2016 June 18th in the framework of the observational programme 097.D-0338. For each spectral range four frames were taken in the ABBA sequence. The spectra were telluric corrected by the reference to a hot B3 standard star HIP\,65630, and by the use of synthetic spectra of the atmosphere in La Silla, generated using the TAPAS\footnote{TAPAS (Transmissions Atmosph\'{e}riques Personnalis\'{e}es Pour l'Astronomie), http://ether.ipsl.jussieu.fr/tapas/, Bertaux \etal (2014)} service.

\begin{figure}
	\centering
	\includegraphics[width=\linewidth]{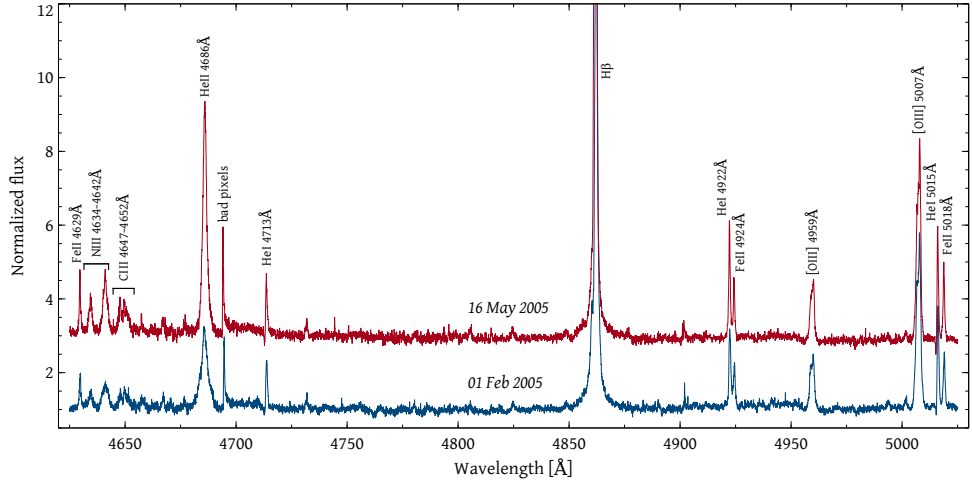}
	\caption{A sample of the FEROS spectra of \st2\ obtained on February 1 and May 16, 2005. The strongest H$\beta$ emission is truncated for clarity.}
\end{figure}

The processing of the SOFI spectra, the reduction of the FEROS spectra into fluxes and all measurements were made with \textsc{iraf}\footnote{\textsc{iraf} is distributed by the National Optical Astronomy Observatories, which are operated by the Association of Universities for Research in Astronomy, Inc., under cooperative agreement with the National Science Foundation.}. The journal of our spectral observations is presented in Table~1.

The fluxes of several emission lines measured by integrating the area under the whole profile are shown in Table~2. Considering the inaccuracies of the reduction, the measurement error for these fluxes is $\sim20$\% and $\sim30$\% for strong and weak lines respectively.

The All Sky Automated Survey (ASAS, Pojmanski 1977) observed \st2\ in $V$ band during the time when our spectra were obtained. Photometry in $I$ band was secured by the Optical Gravitational Lensing Experiment survey (OGLE IV, Udalski \etal 2015) between 2013 May and 2016 June. We use photometry of \st2\ from the Wide-field Infrared Survey Explorer (WISE, Wright \etal 2010). Also, Multiepoch Photometry is available in the AllWISE database for two sets of observations in 2010 February and August. The object was detected also by the Two Micron All Sky Survey (2MASS, Skrutskie \etal 2006). The Yale/San Juan Southern Proper Motion Catalog 4 (SPM4, Girard \etal 2011) gives for \st2\ $B=16\fm92$ and $V=15\fm32$. The existing photometry in different bands is summarized in Table~3.

\MakeTable{lllllllll}{12.5cm}{De-reddened emission line fluxes in units of $10^\mathrm{-13}$\,ergs\,cm$^\mathrm{-2}$\,s$^\mathrm{-1}$.}
{\hline			
\multicolumn{1}{c}{Date} & \multicolumn{1}{c}{JD} & \multicolumn{1}{c}{H$\gamma$} & \multicolumn{1}{c}{[OIII]} & \multicolumn{1}{c}{HeI} & \multicolumn{1}{c}{HeII} & \multicolumn{1}{c}{H$\beta$} & \multicolumn{1}{c}{[OIII]} & \multicolumn{1}{c}{[OIII]} \\
& & \multicolumn{1}{c}{4341\,\AA} & \multicolumn{1}{c}{4363\,\AA} & \multicolumn{1}{c}{ 4471\,\AA} & \multicolumn{1}{c}{4686\,\AA} & \multicolumn{1}{c}{4861\,\AA} & \multicolumn{1}{c}{4959\,\AA} & \multicolumn{1}{c}{5007\,\AA} \\
\hline
2005 Feb 01 & 2453402.252 & 7.82 & 2.18 & 1.64 & 6.98 & 29.5 & 2.90 & 8.62 \\
2005 May 16 & 2453506.077 & 35.3 & 10.0 & 4.90 & 30.0 & 82.4 & 6.74 & 21.4 \\
\hline
}

\MakeTable{lllll}{12.5cm}{Available optical and infrared photometry of \st2.}	
{\hline
Band & $\lambda$ [$\mu$] & Magnitude & Uncertainty & Catalogue \\
\hline
B & 0.44 & 16.92 &  & SPM4 \\ 
V & 0.55 & 15.32 &  & SPM4 \\ 
J & 1.25 & 9.73 & 0.02 & 2MASS \\ 
H & 1.65 & 8.68 & 0.03 & 2MASS \\ 
K$_\mathrm{s}$ & 2.17 & 8.21 & 0.02 & 2MASS \\ 
W1 & 3.4 & 8.13 & 0.02 & WISE \\ 
W2 & 4.6 & 8.20 & 0.02 & WISE \\ 
W3 & 12 & 7.77 & 0.02 & WISE \\ 
W4 & 22 & 7.26 & 0.07 & WISE \\ 	
\hline
}


\section{Results}

\subsection{Bipolar jets ejected during an unnoticed outburst}

During most of the time of the ASAS observations in the $V$ band, \st2\ was below the detection limit.  Only for about five months, between 2005 January and June, the star was brighter than 14\fm6 (Fig.~2), reaching a maximum about 13\fm8 in the beginning of March. If we suppose that the SPM4 value $V=15\fm32$  is close to the quiescence brightness of the star, an amplitude of $\sim1\fm5$ for the 2005 event can be estimated. This is the amplitude, typical for the outbursts of classical symbiotic stars like Z~And and AG~Dra.

The FEROS spectra also support the assumption that an outburst of \st2\ occurred in 2005. In the blue part, a hot continuum fills the M spectrum features and they became well visible around and red-ward of H$\alpha$. The line spectrum is dominated by emissions and the strongest are the Balmer series members. Most numerous among the other emissions are the lines of FeII, HeI, and SiII (Fig.~1). In the near infrared, well visible in emission are the higher members of the Paschen series and the CaII triplet. OI 8446\,\AA\ emission line is remarkably stronger in comparison to  OI 7774\,\AA, suggesting Ly$\beta$ fluorescence excitation. HeII 4686\,\AA\ and the blends NIII 4634--4642\,\AA\ and CIII 4647--4652\,\AA\ (Fig.~1) are present in both spectra while weak emissions of HeII 5412\,\AA\ and [FeVII] 6087\,\AA\ appear in the May 16th spectrum only. The forbidden lines of [OIII] 5007\,\AA, 4959\,\AA\ and 4363\,\AA\ (Fig.~1) are present in the spectrum as well. Weak emissions  [OI] 6300\,\AA, 6364\,\AA\ and [NII] 6584\,\AA\ are also apparent. 

\begin{figure}
	\centering
	\includegraphics[width=\linewidth]{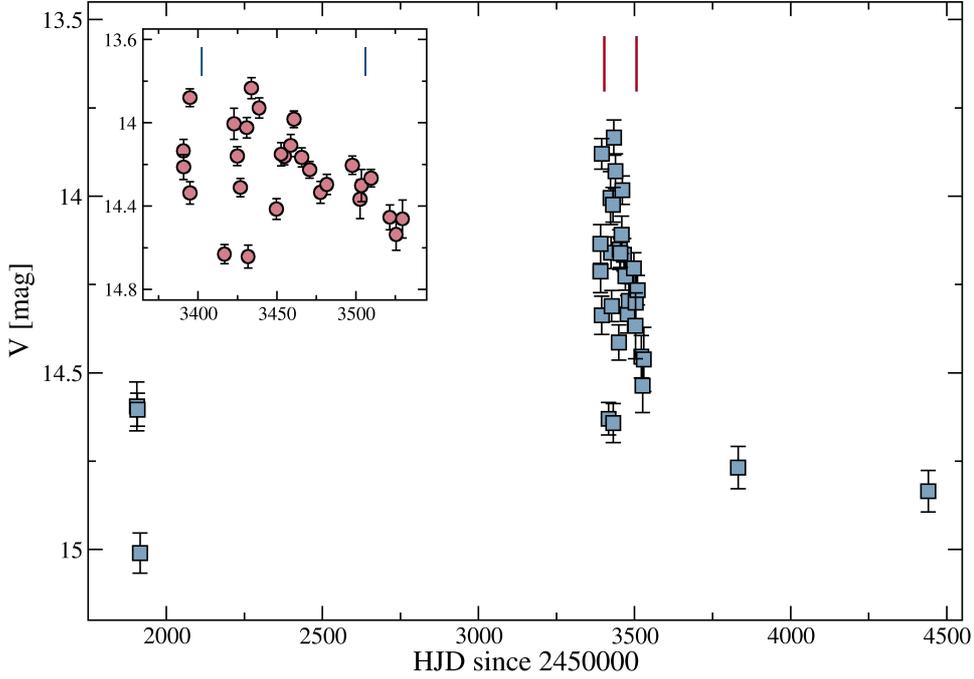}
	\caption{ASAS light curve of \st2\ in $V$ band (\textsl{squares}) and zoomed view around the maximum (\textsl{circles}). The vertical lines mark the moments of our spectral observations.}
\end{figure}

Weak absorptions in the blue emission wings of some HeI lines in the spectrum obtained in February are visible. In the spectrum obtained in May, these absorptions slightly increased in intensity and the profiles became of P~Cyg type. The terminal velocity of the P~Cyg absorption components does not exceed 100\,\kms.

A comparison of H$\alpha$ observed by Van Winckel \etal (1993) with our observations (Fig.~3) shows that the general line shape did not vary to much. The most remarkable change is  the weakening of the absorption component in the 2005 May 16th spectrum. Our measurements also show that there are no significant shifts in the radial velocities. We estimate the average radial velocity for the central emission peak of the four H$\alpha$ lines shown in Fig.~3  to $31.9\pm1.1$\,\kms. Using the absorption lines in the 8400--8850\,\AA\ wavelength range (most probably from the the M giant spectrum), we obtained radial velocities $36.4\pm0.3$\,\kms\ and $35.0\pm0.5$\,\kms, on February 1st and May 16th respectively. Accordingly, the average radial velocities measured for the metallic emission lines, mainly of FeII, on these dates are $29.2\pm0.7$\,\kms\ and $22.3\pm0.4$\,\kms.

A careful inspection of the 2005 spectra revealed two satellite emission components, marked  S$^\mathrm{-}$ and S$^\mathrm{+}$ in Fig.~3, in the emission wings on both sides of the main H$\alpha$ profile. We interpret these satellite components as emissions originating in high-velocity, bipolar outflows, ejected by the hot component of \st2. Such emission components are not visible in the 1988 and 1992 H$\alpha$ profiles (Fig.~3). However, we cannot conclude that they are completely missing because, the region around H$\alpha$, covered by the observations of Van Winckel \etal (1993), is very limited. 

To separate the jet emission components and to estimate their parameters we fitted the complete H$\alpha$ line profiles with a combination of Gaussian and Lorentzian functions. The satellite emissions are best fitted with Gaussian curves, whose parameters are shown in Table~4. Uncertainty in the determination of the parameters is the most significant for the blue jet emission component in the 2005 May 16th spectrum (Fig.~3), as it is blended with the relatively strong line of FeII 6516\,\AA. 

Using the parameters from Table~4 and taking into consideration the shift of the H$\alpha$ central emission peak, the estimated velocities of the bipolar jets are $1555\pm13$\,\kms\ on February 1st and $1847\pm20$\,\kms\ on May 16th. The average FWHM values of the fitted gaussians are $277\pm16$\,\kms\ and $447\pm21$\,\kms, which indicates a high collimation of the outflowing matter. An increase of the jets velocity, by about 300\,\kms, apparently took place in 2005, between February and May.

\begin{figure}
	\centering
	\includegraphics[width=.98\linewidth]{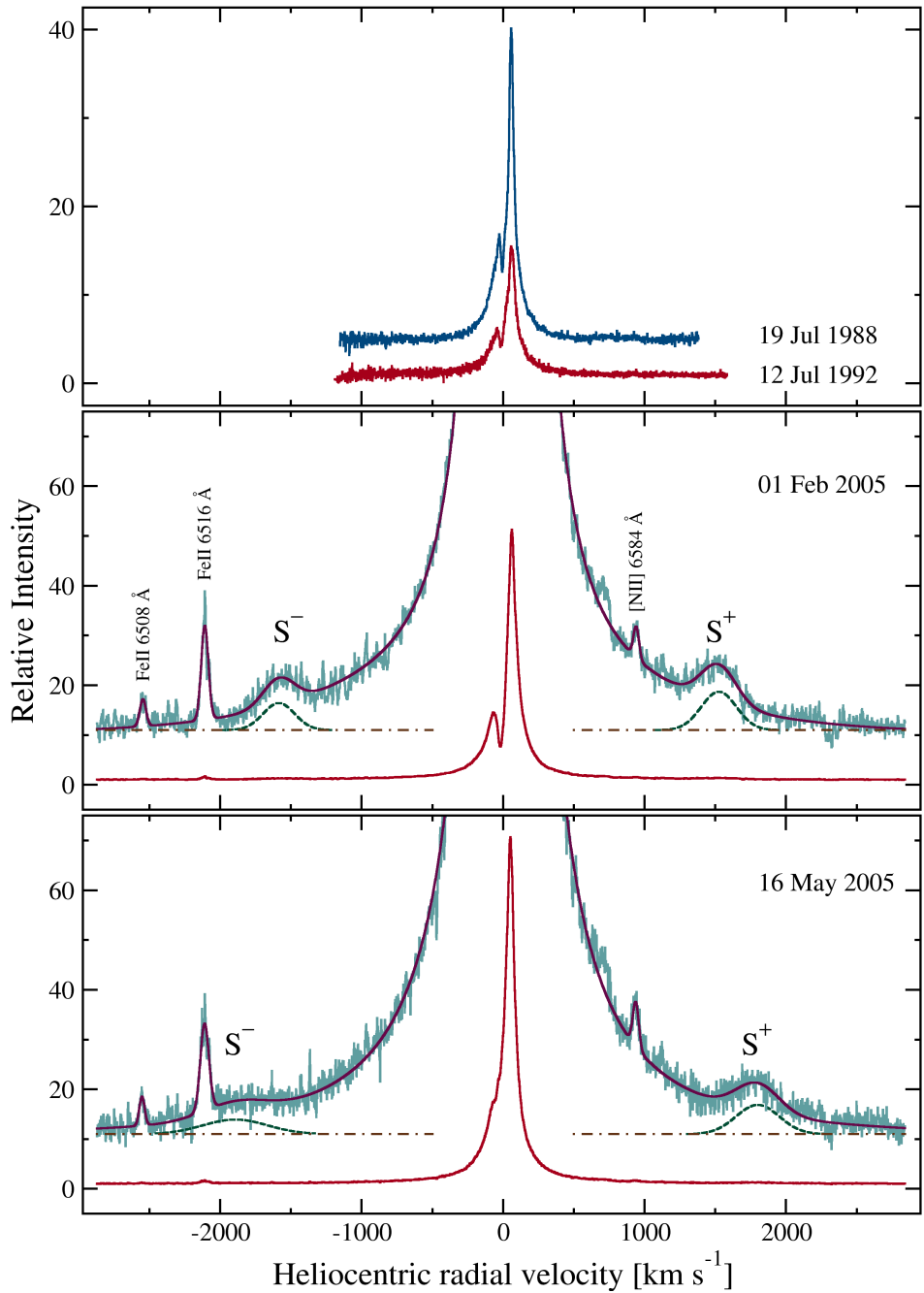}
	\caption{H$\alpha$ profiles in the spectrum of \st2\ obtained by Van Winckel \etal (1993) in 1988 and 1992 (\textsl{upper panel}), compared to our observations in 2005 (\textsl{middle and lower panels}). The profiles are normalized to the local continuum and when necessary shifted for clarity. The 2005 profiles are multiplied by 40 to enlarge their wings and show better the jets emission components S$^\mathrm{-}$ and S$^\mathrm{+}$. The \textsl{dark continuous lines} represent the fit to the enlarged H$\alpha$ profiles (see the text for details). With \textsl{dashed lines} are shown the Gaussian fits to the jet emission components. The \textsl{dot-dashed lines} mark the local continuum level for the enlarged profiles.}
\end{figure}

\MakeTable{lllllllll}{12.5cm}{Heliocentric radial velocities (RV$_{\sun}$), FWHM, and equivalent width (EW$_{\lambda}$) of the gaussian fit to the jet emission components.}
{\hline
\noalign{\smallskip}
\multicolumn{1}{c}{Date} & \multicolumn{2}{c}{RV$_\sun$}\ [\kms] && \multicolumn{2}{c}{FWHM [\kms]} && \multicolumn{2}{c}{EW$_\lambda$\ [\AA]} \\
\cline{2-3}\cline{5-6}\cline{8-9} 
\noalign{\smallskip}
&\multicolumn{1}{c}{S$^{-}$} & \multicolumn{1}{c}{S$^{+}$} && \multicolumn{1}{c}{S$^{-}$} & \multicolumn{1}{c}{S$^{+}$} && \multicolumn{1}{c}{S$^{-}$} & \multicolumn{1}{c}{S$^{+}$} \\
\noalign{\smallskip}
\hline
\noalign{\smallskip}
01.02.2005 & $-1585\pm10$ & $1525\pm8$ &  & $262\pm13$ & $292\pm10$ &  & $0.82$ & $1.30$ \\ 
16.05.2005 & $-1893\pm19$ & $1800\pm6$ &  & $535\pm19$ & $359\pm8$ &  & $0.81$ & $1.21$ \\ 
\noalign{\smallskip}
\hline
}

\subsection{Reddening and distance}

The mentioned above values, $E(B-V)\sim1^\mathrm{m}$ and $d\sim5$\,kpc, reported by Miko\l{}ajewska \etal (1997), are the only estimates of the reddening and distance to \st2\ existing in the literature. The NASA/IPAC Infrared Science Archive gives $E(B-V)=0\fm80\pm0\fm03$ and $E(B-V)=0\fm93\pm0\fm04$ for the mean color excess in the direction of \st2, in accordance with Schlafly \& Finkbeiner (2011) and Schlegel \etal (1998) respectively.

The NaI and KI interstellar lines in our spectra cannot be used to estimate interstellar extinction, because they are heavily blended. Using the equivalent widths of the DIBs  5780\,\AA, 5797\,\AA, 6614\,\AA\ and the dependencies given by Puspitarini \etal (2013) we get for the color excess of \st2\
 $E(B-V)\sim0\fm9\pm0\fm5$. 
 
 The contribution of the hot component to the star brightness in $V$ could be significant and because of this we tried to determine the distance modulus of the object based on its brightness in the near IR. With the color excess obtained by us, we have de-reddened the 2MASS magnitudes of \st2\ in the way described by Li \etal (2016). Then, using their empirical fit for the M giants in the Sgr stream core region, for the absolute magnitude of \st2\ in filter $J$ we obtained $M_\mathrm{J}\sim-5^\mathrm{m}$. This value relates to a distance of the order of 7\,kpc. On the basis of the relationship between the $M_\mathrm{K_s}$ and the color $J-K_\mathrm{s}$ (Sheffield \etal 2014), for an M giant with parameters similar to the determined in Section~3.4, we estimated an absolute magnitude $M_\mathrm{K_s}\sim-5\fm4$. The corresponding distance is about 5.2\,kpc.
 
Obviously, we are far from accurate estimates of the reddening and distance to \st2. Therefore, for the purposes of this work, we will adopt the values of Miko\l{}ajewska \etal (1997) $E(B-V)\sim1^\mathrm{m}$ and $d\sim5$\,kpc.

\subsection{Hot component}

We used  the Iijima (1981) method to estimate the hot component temperature. The method is based on the emission line fluxes of HeII 4686\,\AA, H$\beta$ and HeI 4471\,\AA, assuming Case B recombination. The temperatures, calculated with de-reddened fluxes from Table~2 for, 2005 February 1st and May 16th are 115000\,K and 130000\,K respectively, with an accuracy of the order of 20\%.  The lower limit for $T_\mathrm{h}$ can be derived from the maximal observed ionization potential (IP), using the relation proposed by M\"{u}rset \& Nussbaumer (1994) for temperatures below 150000\,K, $T_\mathrm{h}=\mathrm{IP}\times1000$\,[K]. The emissions with the highest IP$\sim$55\,eV in the February spectrum are HeII 4686\,\AA\ and [OIII] 5007\,\AA. In May a weak emission of [FeVII] 6087\,\AA, with IP of about 125\,eV (Kramida \etal 2015) appeared in the spectrum. 

To calculate the luminosity of the hot component from the de-reddened fluxes of HeII 4686\,\AA\ and H$\beta$ we used equations 6 and 7 from Miko\l{}ajewska \etal (1997). To evaluate the number of H$^\mathrm{0}$ and He$^\mathrm{+}$ ionizing photons, we used the number of ionizing photons $G_\mathrm{i}(T_\mathrm{*})$ tabulated by Nussbaumer \& Vogel (1987).  The difference between the luminosity calculated based on the HeII 4686\,\AA\ and H$\beta$ does not exceed 20\%. As a result, we adopted the average of the values obtained from both equations for the hot component luminosity. This gives an estimate of $L_\mathrm{h}\sim285\pm30$\,L$_\sun$ for the \st2\ on 2005 February 1st and  $L_\mathrm{h}\sim940\pm200$\,L$_\sun$ on May 16th. Considering the uncertainties in the reddening and the distance, the error in the luminosity can increase by a factor of 2. From the values of the temperature and the luminosity, we can evaluate the radius of the pseudo photosphere of the hot component to $ R_\mathrm{h}=0.04\pm0.01$\,R$_\sun$ and $R_\mathrm{h}=0.06\pm0.01$\,R$_\sun$ on 2005 February 1 and May 16 respectively.

The relatively small ratios of [OIII] 5007\,\AA\ to H$\beta$ ($0.27\pm0.02$) and [OIII] 4363\,\AA\ to H$\gamma$ ($0.29\pm0.01$) indicate a comparatively high electron density $N_\mathrm{e}$, for the environment in which the forbidden lines originate. Using the task \textsc{temden} of the \textsc{iraf} package \textsc{stsdas} and adopting a value of $N_\mathrm{e}\sim10^\mathrm{7}$ we obtained the electron temperature of $T_\mathrm{e}\sim11000\,K$ on February 1st and $T_\mathrm{e}\sim14000\,K$ on May 16th.

\subsection{Cool component}

\smallskip
\textsl{3.3.1 Photometry}
\medskip

\begin{figure}
	\centering
	\includegraphics[width=\linewidth]{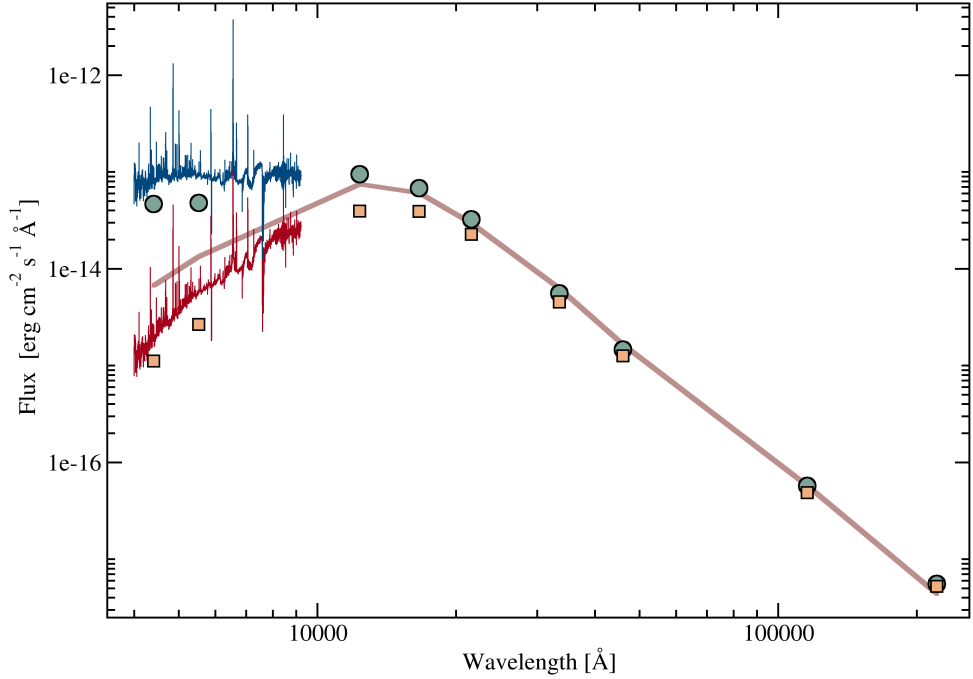}
	\caption{\st2\ SED based on SPM4 $B$ and $V$, 2MASS $J, H, K_\mathrm{s}$, WISE $W1, W2, W3$ and $W4$ magnitudes. The \textsl{squares} indicate the observed values. The de-reddened magnitudes are shown with \textsl{circles}. Observed and de-reddened optical spectrum of \st2, obtained on 2005 February 1st, is also plotted in the figure. The thick \textsl{continuous line} represents the NextGen theoretical spectrum SED (see text for details).}
\end{figure}

To evaluate the physical parameters of the cool component in \st2\ we used the 2MASS $JHK_\mathrm{s}$ data. First, they were transferred to the homogenized system of Bessell \& Brett (1988) (BB) and then de-reddened with the corresponding $A_\lambda$, calculated for $E(B-V)=1\fm0$. The bolometric correction $BC_\mathrm{K}=2\fm78$ for the color $(J-K)_\mathrm{BB}=1\fm06$ was estimated from the respective equation in the paper of Bessell \& Wood (1984).  Using the de-reddened value $K_\mathrm{BB}=7\fm81$ and $BC_\mathrm{K}$ we derived a bolometric absolute magnitude for the red giant $M_\mathrm{bol}=-2\fm9$, which corresponds to the bolometric luminosity $L_\mathrm{bol}=1140$\,M$_\sun$. Using the above value of $(J-K)_\mathrm{BB}$ and $(H-K)_\mathrm{BB}=0\fm25$, and assuming [Fe/H]$=0$ and $\mathrm{log} g=1$,  we evaluated, from Worthey \& Lee (2011), the cool component temperature as $T_\mathrm{eff}=3580\pm100$\,K. We have also calculated the giant radius $R_\mathrm{g}=90$\,R$_\sun$. A comparison of the obtained physical parameters with the calibrations of Straizys \& Kuriliene (1981), points to a M3-M4 red giant.

We used the VO Sed Analyzer (VOSA)  tool (Bayo \etal 2008) to study the spectral energy distribution (SED) of \st2, shown in Fig.~4. The same value of $E(B-V)=1\fm0$ and the extinction law by Fitzpatrick (1999), improved in the infrared by Indebetouw \etal (2005), were used for de-reddening of the observed magnitudes. The disagreement between the optical brightness and the optical spectrum obtained around the 2005 outburst maximum, indicates that the $BV$ magnitudes were measured during quiescence. The IR magnitudes were fitted by a NextGen theoretical spectrum (Allard \etal 2012) with $T_\mathrm{eff}=3500$\,K,  $\mathrm{log} g=1.0$, [Fe/H]$=0$, with the corresponding $L_\mathrm{bol}=1017\pm4$\,L$_\sun$, and using the assumed distance of 5\,kpc. The fitted model demonstrates a good accordance between the IR SED, the M4.5 giant proposed by M\"{u}rset \& Schmid (1999) and the parameters estimated above for the cool component of \st2. Also, the lack of IR excess in Fig.~4 is obvious.

The OGLE data includes only measurements of the $I$ brightness of \st2, obtained at 140 epochs during a time interval of 1120 days (Fig.~5). The observations are scarce, unevenly distributed and with large gaps in between. It should be noted, that because the final calibration to the OGLE-IV Galactic disk photometry is yet to be done, an offset of the zero points of the $I$-band photometry by $\sim$0\fm4 is still present. The full range of the changes in the brightness, with different characteristic times, is $\sim0\fm3$. A gradual increase of the brightness by about $0\fm1$ is also seen in Fig~5. The periodogram analysis of the complete $I$ light curve of \st2, carried out using the phase dispersion minimization method (Stellingwerf 1978) and Period04 (Lenz \& Breger 2005) program, did not show any significant periodicities of the brightness variations. While analyzing only the part of the light-curve between JD\,2456693 and JD\,2456850 covered best with observations, a period of $51\pm7$ days becomes significant. Such a period is most likely caused by pulsations of the cool component in the system, and is in good agreement with the minimal pulsation periods of red giants in symbiotic stars published by Gromadzki \etal (2013) and Angeloni \etal (2014). However, additional observations and more detailed analysis of the pulsations are needed for a definitive conclusion whether the M giant in \st2\ belongs to the OGLE small amplitude red giants (OSARG) or to the semi-regular variables (SRV). For a discussion on OSARG, SRV and symbiotic red giants see Gromadzki \etal (2013) and Angeloni \etal (2014). The deep minimum at JD\,2456814 is most probably due to a superposition of the 51 days pulsations and a decrease in the brightness because of changes with a longer period.

\begin{figure}
	\centering
	\includegraphics[width=\linewidth]{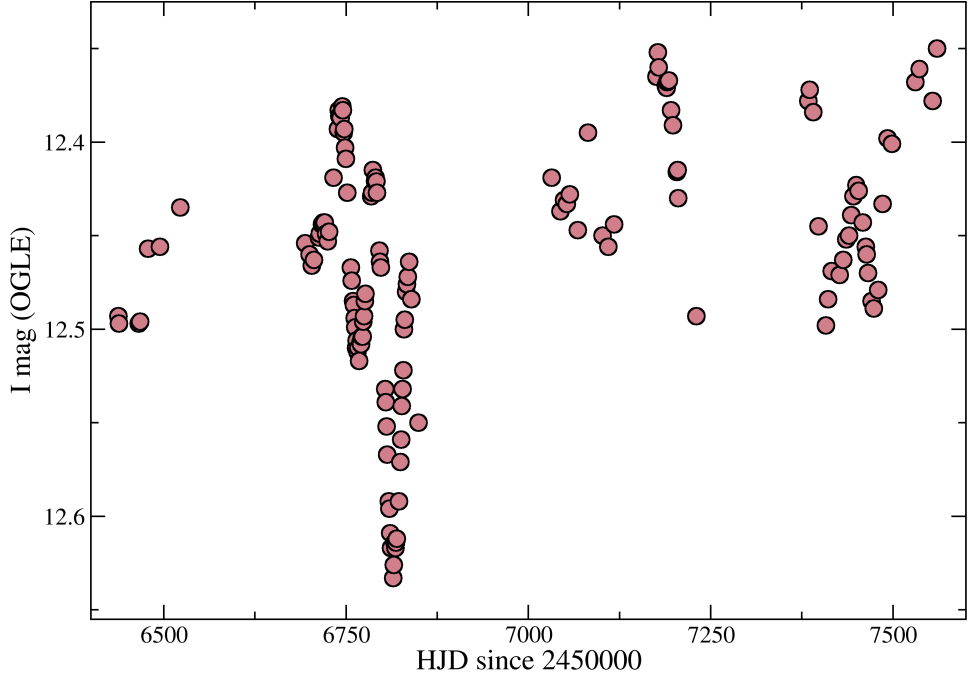}
	\caption{OGLE $I$-band light curve of \st2, covering the period between 2013 May and 2016 June. The error of all measurements is 0\fm003.}
\end{figure}

The multi-epoch AllWISE photometry (Fig.~6) also evidences small changes in the IR brightness of the M giant in the \st2\ system. In the first two WISE bands $W1$ and $W2$ the magnitude values in 2010 February are visibly below the average brightness, marked by dashed lines in Fig.~6. Vice versa, these values in 2010 August are evidently above the average brightness. These differences are of the order of $0\fm05$ and $0\fm03$ for $W1$ and $W2$ respectively. In the other two bands $W3$ and $W4$, measured magnitudes are more or less equally distributed around the average values during both periods of observation.  

\bigskip
\textsl{3.3.2 Near infrared spectroscopy}
\medskip

In the SOFI blue spectra emission lines HeI 10838\,\AA, OI 11292\,\AA, Pa$\beta$ 12823\,\AA\  are clearly seen, indicating that the blue part of the spectrum is substantially affected by the nebular continuum, while emission lines form helium and the hydrogen Bracket series are absent in the red part. We used the parts of $H$ (15310 -- 17450\,\AA) and $K$ (20250 -- 24600\,\AA) band regions, which are poorly affected by the nebula and where the absorption features are relatively strong, to estimate the stellar parameters, to analyse the metallicity and to obtain information on the relations between C, N, and O abundances, and the ratio of $^\mathrm{12}$C/$^\mathrm{13}$C.

Synthetic spectra were calculated using a grid of MARCS model atmospheres (Gustafsson \etal 2008) with the following atmospheric parameters - $T_\mathrm{eff}$ from 3400\,K to 3800\,K, $\mathrm{log} g$ from $+0.5$ to $+1.5$, metallicity [Fe/H] from $+1$ to $-4$. The best fitted solutions were obtained for [Fe/H] $\sim -0.25$. The value, in accordance with the typical for red giants in the S-type symbiotic systems metallicity, between $0.0$ and $-0.5$ (Ga\l{}an \etal 2016, 2017). $\mathrm{log} g$ was between $+1.0$ and $+0.5$, with preference of the higher value. The dependence of the temperature is very weak and the best results are placed around $\sim 3600 - 3700$\,K. The errors are difficult to estimate, but because of the continuum problem (and thus degeneracy especially strong in $T_\mathrm{eff}$) the accuracy is not better than $\Delta T_\mathrm{eff} \sim 200$\,K, $\Delta \mathrm{log} g \sim 0.5$, $\Delta \mathrm{[Fe/H]} \sim 0.5$. In the spectra there are $^\mathrm{12}$CO and $^\mathrm{13}$CO bands after 22900\,\AA\ (Fig.~7) which enable us to measure the abundance of carbon C $\sim 7.8$\,dex (using the model value O $\sim8.4$\,dex), the ratio of C/O $\sim 0.3$, and the carbon isotopic ratio $^\mathrm{12}$C/$^\mathrm{13}$C $\sim 15$. Also weak CN lines are present in both $H$ and $K$ band regions and from them we can roughly estimate the nitrogen abundance N $\sim 8.3$\,dex and thus the ratios of C/N $\sim 0.3$, and O/N $\sim 1.1$. The obtained values indicate that the giant has experienced the first dredge-up, common to all giants in the S-type symbiotic systems studied so far for abundances (Ga\l{}an \etal 2016, 2017). An increased abundance of Sc in the M giant also seems possible when comparing synthetic and observed spectra of \st2.

\section{Discussion}

Our study confirms the classification of \st2\ as a S-type symbiotic star in the catalogue of Belczy\'{n}ski \etal (2000). In fact, on the [OIII]\,5007\,\AA/H$\beta$ versus [OIII]\,4363\,\AA/H$\gamma$ diagnostic diagram by Gutierrez-Moreno \etal (1995), the measured line ratios (Section~3.3) place \st2\ in the same region as the S-type symbiotics. The observed colors $J-H=1\fm06$ and $H-K_\mathrm{s}=0\fm47$ also place \st2\ exactly among the S-type symbiotic stars in the 2MASS color--color diagram, used by Corradi \etal (2008), in combination with the INT Photometric H$\alpha$ survey of the Northern Galactic plane (IPHAS), to distinguish symbiotic binaries from other types of objects. The determined by us $M_\mathrm{bol}=-2.9$\,M$_\sun$, $R_\mathrm{g}=90$\,R$_\sun$ and $T_\mathrm{eff}=3580$\,K put the cool star in \st2\ on the evolutionary track for the red giants with mass around $1.5$\,M$_\sun$ on Fig.~3 in Miko\l{}ajewska (2007), on which symbiotic giants in the HR diagram are represented.

Using low resolution spectra, obtained between 1984 and 1990, Miko\l{}ajewska \etal (1997) estimated the luminosity of the \st2\ hot component to about 600\,L$_\odot$. This coincides with the mean value of our estimates of the luminosity during the outburst in 2005, obtained in Section~3.3. In our spectra Raman scattered OVI emission lines at 6825\,\AA\ and 7082\,\AA\ are not present. Miko\l{}ajewska \etal (1997) pay particular attention to the Raman scattered lines and found an apparent correlation of the symbiotic stars hot component luminosity with the flux of the 6825\,\AA\ emission. In their Table~1, flux measured for this line in the spectrum of \st2\ is missing which suggests that the Raman scattered emission was also absent during their observations. On the other hand, Allen (1984) pointed out the presence of a strong 6825\,\AA\ emission in the spectrum of \st2. Taking into account that the Raman scattered features typically disappear from the spectra of the classical symbiotics during outburst (Tomov \etal 2000; Skopal \etal 2009; Shore \etal 2010), we can suppose that the spectra used by Miko\l{}ajewska \etal (1997) have been obtained during one previous, also unnoticed, outburst of \st2.

\begin{figure}
	\centering
	\includegraphics[width=\linewidth]{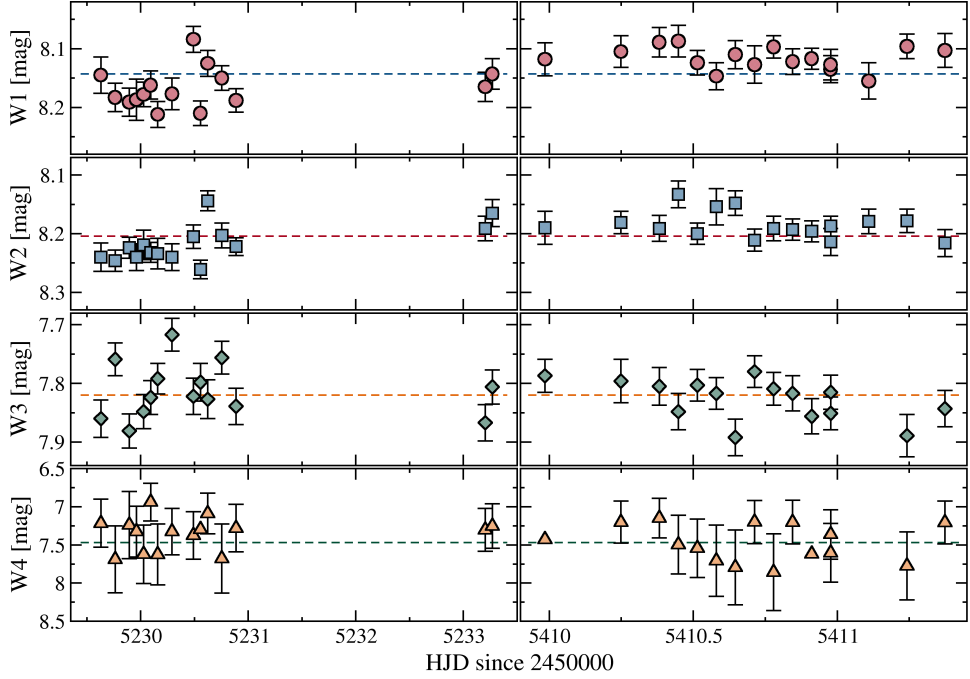}
	\caption{AllWISE multi-epoch photometry light curves of \st2. In the left and the right panels the observations in 2010 February and August are respectively shown. The \textsl{dashed lines} mark the average magnitude of all WISE observations in the corresponding band.}
\end{figure}

As we mentioned above, it is commonly accepted that in all systems producing collimated jets, there is an accretion disk around the central object and the jet velocity is of the order of the escape velocity from this object. To get an idea about the possible angles of the \st2\ orbital inclination $i$ we assume that the jets are ejected perpendicularly to the accretion disk and that the disk lies in the orbital plane. As the minimum and the maximum for the mass of the WD in \st2\ system we use the values 0.4\,M$_\sun$ and 0.8\,M$_\sun$, respectively, adapted from Miko\l{}ajewska (2003). As the minimum radius of the WD we use the typical value 0.01\,R$_\sun$, and as the maximum radius the mean radius of the pseudo photosphere on both dates $R_\mathrm{h}=0.05\pm0.01$\,R$ _\sun$ estimated by us in Section~3.3. As jet's velocity we use the mean value $1700\pm25$\,\kms\ of the velocities obtained for 2005 February 1st and May 16th (Section~3.1). Thus, we find that the inclination of the \st2\ orbit can lie in a wide range between 13\degr\ and 72\degr. 

Skopal \etal (2009) proposed  a formula for estimating the jet's opening angle $\varTheta_\mathrm{jet}$, based on the FWHM of the emission components originating in the jets, the jet's velocity and the orbital inclination. Applying this formula for \st2, using the FWHM values from Table~4, we obtain a very large, unreal $\varTheta_\mathrm{jet}\sim105$\degr\ for the lower limit of $i\sim13$\degr. The estimated jet's opening angle for the upper limit of $i\sim72$\degr\ is $\varTheta_\mathrm{jet}\sim7$\degr. 

For the jets of Z~And, which are very similar in observed velocity and FWHM of the jet emissions to those in \st2, Skopal \etal (2009) calculated $\varTheta_\mathrm{jet}=6.1$\degr\ for $i=76$\degr. An inclination very close to the defined by us upper limit for the orbit of \st2. Another indication that the orbital inclination of \st2\ must be considerable, could be a comparison with MWC~560. In this system, the axis of the jets is parallel to the line of sigh. In the spectrum only the blue jet is visible and in absorption at that (Tomov \etal 1990). Schmid \etal (2001) estimated the orbital inclination of MWC~560 to $i<16$\degr. Similarly, for inclination angles close to 13\degr\ the red jet in \st2\ would be totally obscured by the accretion disk and the blue jet would be visible in absorption. In our spectra, the jets are seen as strong shifted emission components (Fig.~3). Moreover, it seems that the red component is slightly stronger than the blue one, $EW_\mathrm{S^\mathrm{-}}/EW_\mathrm{S^\mathrm{+}}<1$.

Taking into account the above-mentioned facts, we can assume that the inclination of the orbit of \st2\ is large and very likely to be close to the determined by us upper limit of about 72\degr.

It is unclear, where does the mentioned before difference of $\sim300$\,\kms\ between the velocity of the jets in February and May 2005  comes from. It is difficult to explain this on the basis of the jets ejection mechanism, which has not yet been clarified. From the observational point of view, two types of jets velocity changes in symbiotic stars have been observed so far. In some cases, the velocity of the jets gradually changes within 200-300\,\kms in a time scale of several months, which was best observed in Z~And (Skopal \etal 2009).  While the changes in the MWC~560 jets velocity, during the discrete jet ejections in 1990, reached from several hundred to several thousand \kms\ in spectra obtained in two consecutive nights (Tomov \etal 1992). The jets in \st2\ were observed only on two occasions, separated by two and a half months. This does not allow us to trace in detail their evolution and the way of the change of their velocity.

Zamanov \etal (2008) estimated the projected rotational velocity of the cool component in \st2\ as $v\,\mathrm{sin}\,i=9.8\pm1.5$\,\kms. Assuming that the rotational axis of the M giant is perpendicular to the orbital plane, and using the determined upper limit of 72$\degr$ for the orbital inclination, we obtain the rotational period of the giant of $P_\mathrm{rot}\sim445$ days. There are suggestions that the rotational period of an M giant and the orbital period in the S-type symbiotics are synchronized (see for instance Zamanov \etal 2007). Assuming that the synchronization is also occurring in the case of \st2\, the derived value $P_\mathrm{orb}\sim445$ days is in good agreement with the known orbital periods of the S-type symbiotic stars (see Table~1 in Miko\l{}ajewska 2003). Using 1.5\,M$_\sun$ and 0.5\,M$_\sun$ as the masses of the red giant and the white dwarf respectively, from the third Kepler's law we obtain for the binary semi-major axis the value $a\sim307$\,R$_\sun$. This gives $a\,\mathrm{cos}\,i\sim95$\,R$_\sun$, a value greater but close to the sum of the M giant radius $\sim90$\,R$_\sun$ and the radius of the pseudo photosphere of the hot component $\sim0.05$\,R$_\sun$. Taking into account the accuracy of the estimated stellar and orbital parameters, and the fact that we do not consider the accretion disk, which radius is surely greater  than $R_\mathrm{h}$, we can conclude that eclipses in \st2\ are very likely  for $i\sim70\degr$. Presumably, it will be easier to detect a possible eclipse of the hot component by the red giant in the $UBV$ filters, in which the hot component radiation dominates or at least is substantial. 

\begin{figure}
	\centering
	\includegraphics[width=\linewidth]{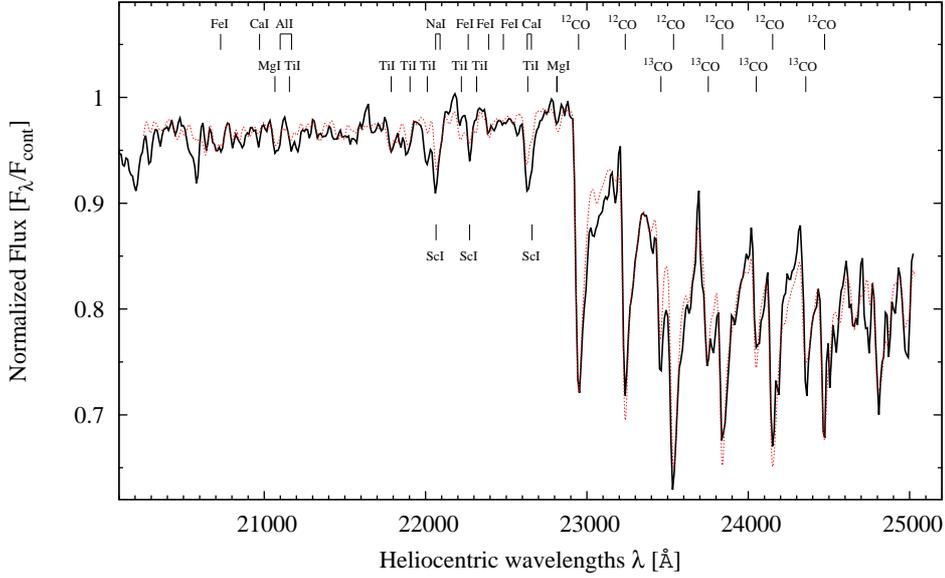}
	\caption{Synthetic spectrum (\textsl{doted line}) generated by the use of MARCS model atmosphere ($T_\mathrm{eff} = 3600$\,K, $\mathrm{log} g = 1.0$, $z = -0.5$) compared to the observed spectrum (\textsl{continuous line}) of \st2. Apart from the CO bands the positions of some of the strongest neutral atomic lines of Na, Mg, Al, Ca, Sc, Ti, and Fe are marked in the figure.}
\end{figure}


\section{Conclusions}

The main results in this paper can be summarized as follows:
\begin{itemize}
\item[(i)] We bring to light a recorded but unnoticed outburst of \st2\ that occurred in 2005. Its amplitude of $\sim1\fm5$ and duration of about half a year, resemble the typical classical symbiotic outbursts.
\item[(ii)] Our study confirms that \st2\ is a S-type symbiotic system. The parameters of the cool component evaluated by us agree, within a subclass, with the spectral class M4.5III proposed by M\"{u}rset \& Schmid (1999). Indications of a possible pulsation period of about 51 days were found in the OGLE light curve.
\item[(iii)] The estimated temperature and the luminosity for the hot component during the 2005 outburst are similar to that obtained by Miko\l{}ajewska \etal (1997). This result and the lack of the Raman scattered emissions in the spectrum of \st2\ in both occasions can be considered as an indication that the spectra used in the paper of Miko\l{}ajewska \etal (1997) were also obtained during a previous, unnoticed outburst.
\item[(iv)] H$\alpha$ satellite emission components, originating in high-velocity, collimated bipolar jets, were identified in the outburst spectra of \st2. Therefore, this poorly studied star, should be added as a new member of the jet-producing group of symbiotic systems.
\item[(v)] On the basis of the only two spectra of \st2\ obtained during the outburst we estimated an average velocity of the jets of about 1700\,\kms. Most likely, the orbital inclination is large, and close to the defined upper limit of about 72$^\circ$. In which case, the opening angle, in good accordance with the profiles of the satellite emissions, will be of the order of 7$^\circ$.
\end{itemize}

Here we tried to present all the accessible, very scarce, observational data for \st2. Additional new or archival data, if available, are necessary, to better understand the nature of this interesting, but until now neglected symbiotic star.

\Acknow{CG has been financed by the Polish National Science Centre grants FUGA No. DEC-2013/08/S/ST9/00581 and SONATA No. DEC-2015/19/D/ST9/02974. This study is based on observations collected at the European Organisation for Astronomical Research in the Southern Hemisphere under ESO programmes 074.D-0114 and 097.D-0338. The research has made use of the NASA's Astrophysics Data System, and the SIMBAD astronomical data base, operated by CDS at Strasbourg, France.This publication makes use of data products from the Wide-field Infrared Survey  Explorer, which is a joint project of the University of California, Los Angeles, and  the Jet Propulsion Laboratory/California Institute of Technology, and NEOWISE, which is  a project of the Jet Propulsion Laboratory/California Institute of Technology. WISE  and NEOWISE are funded by the National Aeronautics and Space Administration. This work makes use of VOSA, developed under the Spanish Virtual Observatory project supported from the Spanish MICINN through grant AyA2011-24052. We are grateful to the anonymous referee for valuable comments and suggestions and also to Sz. Zywica for his help with English.}

\end{document}